# Van der Waals phonon polariton microstructures for configurable infrared electromagnetic field localizations

*Wuchao Huang, Fengsheng Sun, Zebo Zheng, Thomas G. Folland, Xuexian Chen, Huizhen Liao, Ningsheng Xu, Joshua D. Caldwell, Huanjun Chen\* & Shaozhi Deng\**

W. Huang, F. Sun, Dr. Z. Zheng, X. Chen, H. Liao, Prof. N. Xu, Prof. H. Chen, Prof. S. Deng
State Key Laboratory of Optoelectronic Materials and Technologies, Guangdong Province Key Laboratory of Display Material and Technology, School of Electronics and Information Technology
Sun Yat-sen University, 510275, China
E-mail: chenhj8@mail.sysu.edu.cn; stsdsz@mail.sysu.edu.cn.
Dr. T. G. Folland, Prof. J. D. Caldwell
Department of Mechanical Engineering
Vanderbilt University
Nashville, TN 37235, USA

Polar van der Waals (vdW) crystals that support phonon polaritons have recently attracted much attention because they can confine infrared and terahertz (THz) light to deeply subwavelength dimensions, allowing for the guiding and manipulation of light at the nanoscale. The practical applications of these crystals in devices rely strongly on deterministic engineering of their spatially localized electromagnetic field distributions, which has remained challenging. This study demonstrates that polariton interference can be enhanced and tailored by patterning the vdW crystal α-$MoO_3$ into microstructures that support highly in-plane anisotropic phonon polaritons. The orientation of the polaritonic in-plane isofrequency curve relative to the microstructure edges is a critical parameter governing the polariton interference, rendering the configuration of infrared electromagnetic field localizations by enabling the tuning of the microstructure size and shape and the excitation frequency. Thus, our study presents an effective rationale for engineering infrared light flow in planar photonic devices.

## 1. Introduction

Phonon polaritons (PhPs) are quasi-particles arising from the coupling of photons and optical phonons in polar crystals.[1–2] Compared to bulk crystals, the recently identified two-dimensional (2D) vdW atomic crystals are attracting increasing attention due to the exotic PhPs they can support, which originate from their highly anisotropic, polar crystal structure.[3–9] Due to in-plane covalent bonding and out-of-plane vdW coupling, the permittivity tensors of 2D vdW crystals can exhibit opposite signs for the components parallel and perpendicular to the stacked planes, making the PhPs hyperbolic in nature. These hyperbolic PhPs (HPhPs) exhibit low losses and high quality factors, which enable the confinement and manipulation of electromagnetic fields at a deeply subwavelength scale and give rise to various important applications such as super-resolution imaging,[10, 11] ultrasensitive sensing,[12] and thermal management.[13, 14] In particular, recently explored 2D transition metal oxides, such as α-MoO$_3$[8, 9, 15, 16] and α-V$_2$O$_5$,[17] represent natural biaxial hyperbolic crystals. The permittivities are highly anisotropic and even exhibit components with opposite signs along orthogonal in-plane directions over a broad spectral range, giving rise to PhPs with in-plane hyperbolicity. These HPhPs are observed as propagating wavefronts of concave shape, in striking contrast to those propagating in uniaxial crystals, such as hexagonal boron nitride (h-BN).[3, 4] Such anisotropic characteristics open up new opportunities for confining and configuring electromagnetic waves at the nanoscale, especially in the mid-infrared to terahertz spectral range with limited photonic material candidates. For example, very recent studies have demonstrated that the stacking and twisting of two α-MoO$_3$ layers result in intriguing planar light propagation and localization behaviors.[18–21]

The practical utilizations of vdW HPhPs strongly rely on the configuration of their electromagnetic fields to overlap with the functional components of a device or photonic circuits. Several approaches have been proposed to engineer HPhPs, including controlling the dielectric environment,[22–25] carrier concentration,[26, 27] material composition,[28, 29] and geometry[30–32] of the crystals and devices. In particular, patterning the crystals into micro- or nanostructures can generate localized electromagnetic fields with configurable spatial distributions that are strongly dependent on geometric parameters.[30–32] Recent studies have successfully demonstrated rich and intriguing localized electromagnetic modes in micro- or nanostructured polaritonic crystals, such as graphene,[33] h-BN,[30–32] SiC,[34] and hybrids.[32, 35] Most of the previous results focused on uniaxial crystals, where the in-plane polariton propagation is isotropic. Recently, subwavelength mid-infrared electromagnetic field localizations have been reported in circular and square α-MoO$_3$ micro-disks.[8, 9] Nevertheless, configuring the near-field distributions associated with the in-plane anisotropic, and critically hyperbolic polaritons, via patterning and the associated design principles is still incomplete. As such developing these understandings of highly anisotropic polariton propagation and spatial confinement is anticipated to open a new degrees of freedom for sub-diffraction-limit light focusing and manipulation.

Here, through the combination of theoretical calculations and real-space nano-imaging, we demonstrate the engineering of infrared electromagnetic field localization of HPhPs in microstructures constructed from biaxial α-MoO$_3$ crystal. The polariton waves that propagate inside a specific microstructure will encounter and be reflected by boundaries, whereby polariton interference results, generating complex standing wave patterns. We show that such interference is governed by the orientation of the in-plane polaritonic isofrequency curve (IFC) relative to the microstructure edges. This is possible because the energy flow directions of the incident and reflected polariton waves are both normal to the IFC. Therefore, the electromagnetic field spatial distributions within a microstructure can be configured by tuning the size and shape, as well as the excitation frequency. The obtained localized fields are highly anisotropic, irrespective of the symmetric microstructure shapes that are supported within. Moreover, we show that the spatial near-field distributions of a wedge-shape microstructure are strongly dependent on the angles between the bisector of the vertex angle and the [001] axis of the α-MoO$_3$ crystal. Such result further illustrates the combination of in-plane anisotropy and boundary manipulation as a paradigm for configuring electromagnetic fields at the nanoscale.

## 2. Results

α-MoO$_3$ supports anisotropic PhPs throughout the mid-infrared (545 to 1010 cm$^{-1}$) and THz (267 to 400 cm$^{-1}$) spectral range,[8, 9, 15, 16] over much of which these modes are hyperbolic in nature. These hyperbolic spectral bands originate from the opposite signed permittivities along different crystalline directions: Re($\varepsilon_i$)·Re($\varepsilon_j$) < 0, with *i* and *j* representing the [100] (*x*-axis), [001] (*y*-axis), and [010] (*z*-axis) crystalline axes (Figure S1a and S1b).[8, 9] In the *x*−*y* plane, the HPhPs exhibit open hyperbolic-shaped IFCs and propagate with concave wavefronts inside Restrahlen Band 1 (545 to 851 cm$^{-1}$) and Band 2 (820 to 972 cm$^{-1}$) (Figure 1a, 1b, 1d, and 1e, Figure S2b and S2c). Even for Restrahlen Band 3 (958 to 1010 cm$^{-1}$) where the IFC is closed due to both in-plane permittivities being positive, the polaritons remain anisotropic featuring elliptical wavefronts (Figure 1g, 1h and Figure S2d) as Re($\varepsilon_x$) ≠ Re($\varepsilon_y$). For ease of discussion, hereafter Band 1, 2, and 3 will be named as negative-$\varepsilon_y$-Band, negative-$\varepsilon_x$-Band, and elliptical band, respectively. Every polariton wave launched into a microstructure will propagate with the energy flow in direction orthogonal to the IFC. This will be reflected by the boundaries of the microstructure and subsequently interfere with the reflected wave, thereby generating standing-waves of various spatially localized electromagnetic field distributions.[33] Because the interference effects originate from the superpositions of the polariton wavefronts, the standing-waves are determined by the wavevector distributions, *i.e.*, polariton IFC topology. Consequently, the anisotropic IFCs can bring a new degree of freedom for configuring the localized electromagnetic fields in α-MoO$_3$ microstructures, which, otherwise is exclusively determined by the microstructure geometry fabricated from crystals with an isotropic IFCs (*i.e.*, Re($\varepsilon_x$) = Re($\varepsilon_y$), Figure S2a, Figure S3a−S3c).[3, 4]

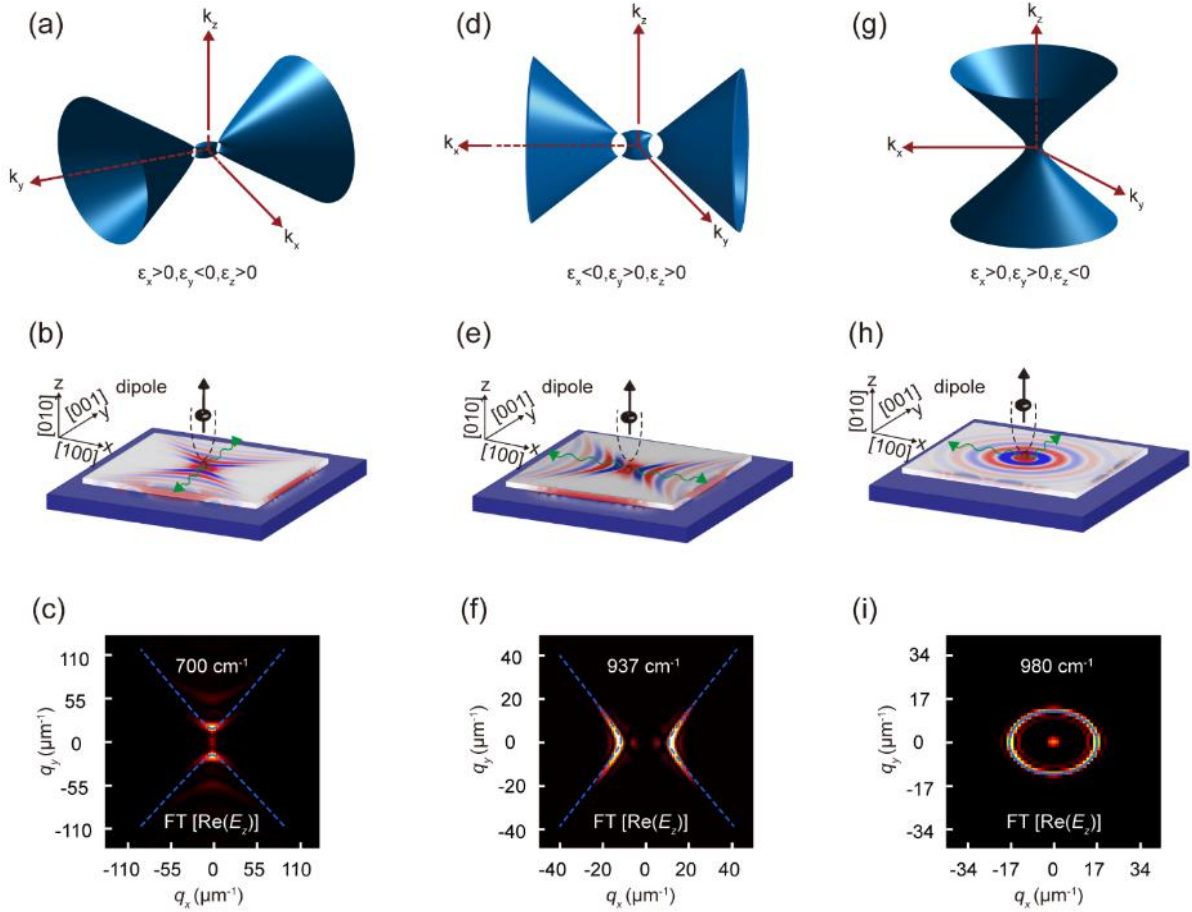

**Figure 1.** HPhPs in biaxial vdW crystals. a, d, and g) Calculated 3D isofrequency contours in the α-MoO$_3$ slab. The calculations were performed at 700 cm$^{-1}$ (negative-$\varepsilon_y$-Band), 937 cm$^{-1}$ (negative-$\varepsilon_x$-Band), and 980 cm$^{-1}$ (elliptical band), respectively. b, e, and h) Calculated Re($E_z$) above the α-MoO$_3$ slab surface at the same three frequencies. Green arrows indicate the propagation directions of the polariton waves. c, f, and i) Isofrequency curves of the α-MoO$_3$ slab at the same three frequencies. The Re($E_z$) distributions were calculated by launching the HPhPs on the sample surfaces using *z*-polarized electric dipoles. The isofrequency curves shown in (c), (f), and (i) were obtained as Fourier transforms of (b), (e), and (h), respectively. The dashed lines represent the in-plane dispersions obtained from the analytical electromagnetic waveguide model calculations. The α-MoO$_3$ slab is 170-nm thick.

An analytical model was first developed to calculate the in-plane IFC $q[\vec{\varepsilon}(\omega), d, \theta]$, of the HPhPs propagating inside a vdW slab of finite thickness (see note S1).[36, 37] To simplify the discussion, $\varepsilon$ denotes the real part of the permittivity in our following discussion. According to the model calculations, the electromagnetic waves in the hyperbolic vdW slab are dominated by the TM polariton modes with a dispersion relation:

$$\sqrt{\frac{\varepsilon_t}{\varepsilon_z}}\sqrt{k_0^2\varepsilon_z - q^2}\,d = \arctan\left(\frac{\sqrt{\varepsilon_t\varepsilon_z}}{\varepsilon_c}\frac{\sqrt{q^2 - k_0^2\varepsilon_c}}{\sqrt{k_0^2\varepsilon_z - q^2}}\right) + \arctan\left(\frac{\sqrt{\varepsilon_t\varepsilon_z}}{\varepsilon_s}\frac{\sqrt{q^2 - k_0^2\varepsilon_s}}{\sqrt{k_0^2\varepsilon_z - q^2}}\right) + M\pi \quad (1)$$

where $\varepsilon_t = \varepsilon_x \cos^2\theta + \varepsilon_y \cos^2\theta$; $d$ is the slab thickness; $k_0$ is the free-space wavevector; $\theta$ denotes the angle of the propagation direction relative to the $x$-axis, $M$ represents the order of the different TM modes, and $\varepsilon_c$ and $\varepsilon_s$ are the dielectric constants for air and the substrate, respectively. Using the permittivity of α-MoO$_3$ as the input parameter (see note S1, table S1), the in-plane IFCs of the slabs were calculated according to Equation (1) at representative frequencies, which agree well with the numerical results (Figure 1c, 1f, and 1i). Specifically, for an α-MoO$_3$ slab, the IFCs in negative-$\varepsilon_x$-Band is hyperbolas opening toward the $x$-axis (Figure 2a). The polariton wave vectors are restricted inside the opening angle of $\varphi = 2\arctan\left(\sqrt{\varepsilon_x(\omega)}/i\sqrt{\varepsilon_y(\omega)}\right)$ bisected by $x$-axis. As a result, polaritons within the Restrahlen band are forbidden to propagate along $y$-axis. Similar calculation results can be obtained for the negative-$\varepsilon_y$-Band (see note S2, Figure S4a). In elliptical band, the IFCs are ellipses (Figure 2b), resulting in polaritons spreading along all directions in the $x$–$y$ plane, but with orientation-dependent wavevectors.

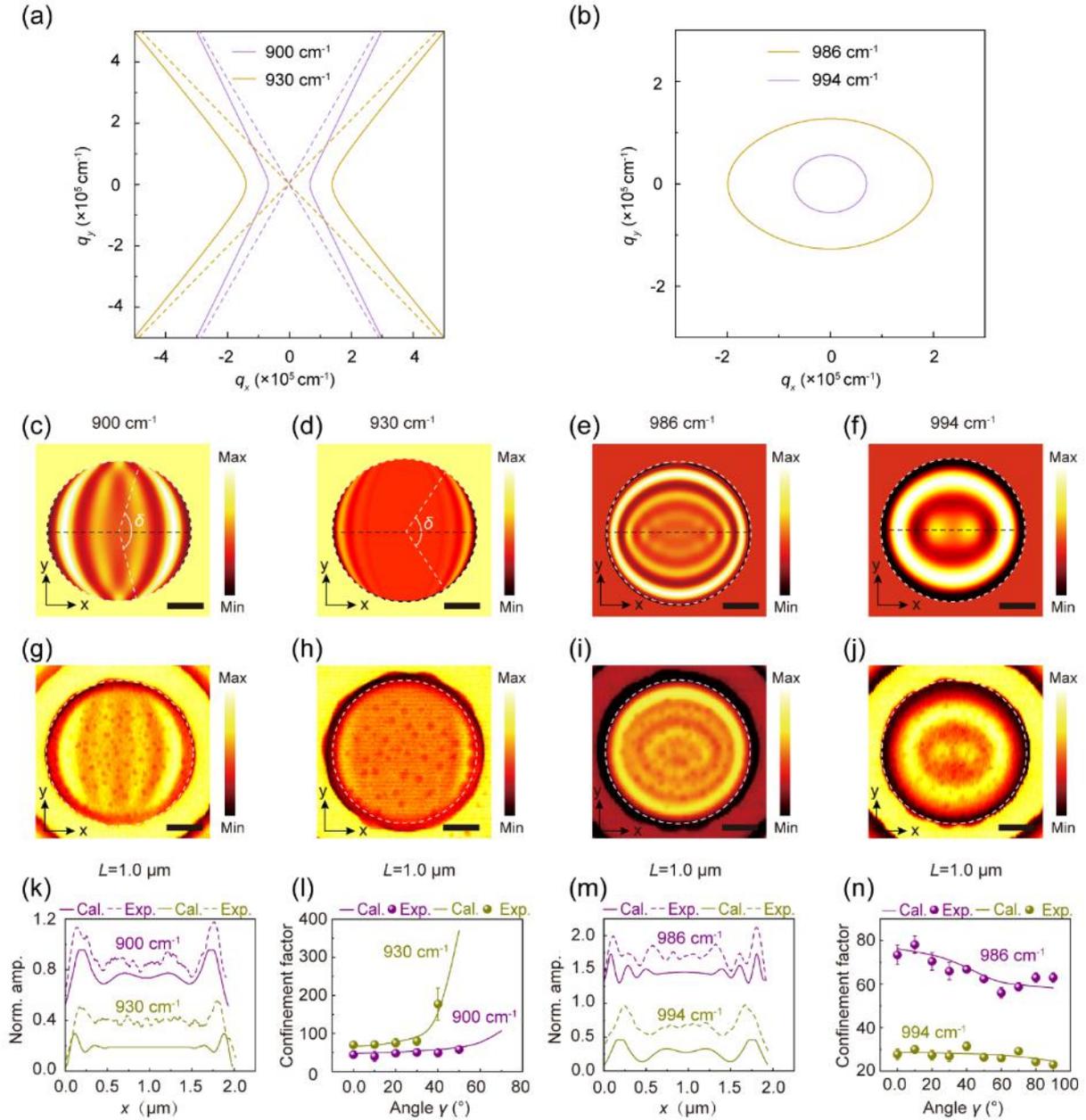

**Figure 2.** HPhP interferences in a α-MoO$_3$ micro-disk. a, b) In-plane IFCs at two representative frequencies in negative-$\varepsilon_x$-Band (a) and elliptical band (b). The dashed lines in (a) denote the asymptotes of the IFCs. c−f) Calculated interference patterns of a 1.0-μm-radius, 170-nm-thick micro-disk at different excitation frequencies using the phenomenological model (see note S3). The two dashed white lines in (c) and (d) indicate the extent of the outmost bright fringes. g−j) Real-space nano-imagings of the corresponding micro-disks. The boundaries are marked with white dashed circles. k, m) Experimental (dashed) and calculated (solid) amplitude profiles extracted along the horizontal dashed lines marked in (c) to (f). The amplitudes are normalized by those from the substrate next to the disk. l, n) Field confinement factors of the micro-disk at different excitation frequencies. The symbols are the experimental values, and the solid lines are the calculation results. The error bars in (l) and (n) are obtained based on the two symmetrical points along the $y$-axis on the outmost fringes. Angle $\gamma$ is defined as the angle with respect to $x$-axis along the center of the outmost fringe. Scale bars: 500 nm.

Having established the IFC, the interference patterns of α-MoO$_3$ microstructures can be readily calculated using the phenomenological cavity model (see note S3).[15, 20, 36, 38] Only the boundary perpendicular to the polariton wavevector is considered in our calculations to simplify the discussion and correlate the model calculations directly with near-field measurements performed using scattering-type scanning near-field optical microscopy (s-SNOM).[8] In other words, the interference effects are dominated by those between a polariton wave and its antiparallel back-reflected wave. As will be shown in the following discussion, this approximation already captures the main physics of the polariton interference in α-MoO$_3$ microstructures. To demonstrate the impacts of IFCs on the interference patterns, we first consider the polariton waves launched into a circular micro-disk of rotational symmetry (radius $L$ = 1.0 μm, thickness $t$ = 170 nm). The circular edge guarantees that all polariton waves propagating through the center can be back-reflected. For two typical frequencies in negative-$\varepsilon_x$-Band, *i.e.*, $\omega$ = 900 cm$^{-1}$ and 930 cm$^{-1}$, the calculation indicates deformed bright (dark) fringes along the *y*-axis due to constructive (destructive) interference that results (Figure 2c and 2d). These features result from reduction of polariton wavelength $\lambda_p$ (corresponding to the fringe spacing) along the *y*-axis, which are consistent with previous results.[8, 9] For the polaritons in elliptical band, the interference fringes are a series of ellipses (Figure 2e and 2f, Figure S5a−S5e).

In addition to the fringe shapes, the anisotropic IFCs can modulate the interference patterns in three aspects. First, in negative-$\varepsilon_x$-Band, the IFCs exhibit a larger $\varphi$ at a smaller $\omega$ (Figure 2a and Figure S4b). Therefore, as $\omega$ is reduced, interference fringes contributed by polaritons with wavevectors close to the IFC asymptotes will extend toward the two disk end points along the *y*-axis (Figure 2c, 2d, and Figure S5f−S5j). Second, in both negative-$\varepsilon_x$-Band and ellptical band, the fringe spacings are smaller for excitation frequencies corresponding to IFCs with larger wavevectors (Figure 2c−2f, 2k, 2m, Figure S5f−S5j, and Figure S5p). As the fringe spacings are reduced, the widths of the bright fringes become smaller, indicating stronger electric field confinements. To quantitatively evaluate the field localization, a confinement factor defined as CF = $\lambda_0/W$ is employed, with $\lambda_0$ and $W$ as the free-space excitation wavelength and full width at half maximum (FWHM) of the outermost bright fringe to the edge, respectively. CF clearly increases as the excitation frequency is increased for negative-$\varepsilon_x$-Band (Figure 2l and Figure S5q), whereas a reverse evolution is observed in elliptical band (Figure 2n and Figure S5r). Third, the field confinements are non-uniform within an individual bright fringe. In negative-$\varepsilon_x$-Band, the CF of the outermost fringe increases as the angle $\gamma$ with respect to $x$-axis becomes larger (Figure 2l and Figure S5q). In contrast, in elliptical band, the CF weakens as $\gamma$ is increased (Figure 2n and Figure S5r). These behaviors can be understood by considering that the evolution of $q$ as a function of $\gamma$ in negative-$\varepsilon_x$-Band is opposite with those in elliptical band (Figure 2a and 2b). It should be noted that polariton interferences behaviors in negative-$\varepsilon_y$-Band are similar to those in negative-$\varepsilon_x$-Band, where the calculated fringes deform along the $x$-axis (see note S2, Figure S5k−S5p, and Figure S5s).

The above model calculation results can be verified experimentally. To that end, α-MoO$_3$ microstructures of various sizes and shapes are fabricated by employing the focus-ion beam (FIB) technique (Figure S6a and Figure S7). The interference patterns are visualized using a real-space nano-imaging approach based on s-SNOM (Figure S6b). As shown in Figure 2g−2j, the measured near-field distributions agree with the calculated interference features, showing similar deformed and elliptical fringes for the excitation frequencies in negative-$\varepsilon_x$-Band and elliptical band. By extracting the respective amplitude profiles crossing the disk center along the $x$- and $y$-axes (Figure 2k, 2m, and Figure S8), it is observed that both numbers of the oscillating maxima and their dependence on the excitation frequency are consistent between the experimental and calculation results. Additionally, the s-SNOM characterizations and the model calculations are in good quantitative agreement in terms of CF dependence on $\gamma$ (Figure 2l and 2n). Note the differences between the experimental and calculated amplitudes close to the disk center, where the nano-images show small fluctuations. We attribute these discrepancies to the defects and impurities introduced during the micro-disk fabrication.

We further corroborate the model calculation results and reveal the complex interference fields by calculating the near-field distributions, Re($E_z$), above the micro-disks launched by a z-polarized electric dipole. Such a simulation configuration guarantees that all reflected waves, and not just the back-reflected ones, are included in the interference fields. The simulations are performed using finite element method (FEM, Comsol). The results indicate that unlike other polaritons generating isotropic concentric patterns,[33, 39] the localized fields in the micro-disk clearly reveal highly anisotropic frequency-dependent spatial distributions. Cross-shape fringes that stem from the directional polariton propagation associated with the in-plane hyperbolic response of the α-MoO$_3$ can be observed near the disk center at $\omega$ = 900 cm$^{-1}$ (negative-$\varepsilon_x$-Band) (Figure 3a and Figure 1e). Meanwhile, elongated concentric fringes that originate from the in-plane elliptical response can be observed for excitation at 994 cm$^{-1}$ (elliptical band) (Figure 3d and Figure 1h). These anisotropic HPhPs will be reflected by the circular edge and generate deformed fringes close to the edge (Figure 3a and 3d). The phase difference between two adjacent bright fringes is approximate π, which is a typical characteristic of a wave interference. Accordingly, the polariton wavelengths can be readily quantified by measuring the separations between the near-field maxima and minima (as indicated in Figure 3a and 3d) yielding $\lambda_p/2$. The extracted polariton wavelengths are consistent with the model calculations and s-SNOM measurements (Figure 3c and 3f). For a more direct comparison with the s-SNOM measurements, near-field images are simulated by recording the $|E_z|$ as a function of the dipole position. The obtained $|E_z(x, y)|$ corroborates with the experimental results in terms of fringe shapes and spacings (Figure 2j, Figure 3b, 3c, 3e, 3f, Figure S8a, and S8b). It should be noted that the polariton field decays faster at 994 cm$^{-1}$ outside the disk than that at 900 cm$^{-1}$ (Figure 3a, 3b, 3d, and 3e). Such a difference can be understood from two aspects. First, polaritons excited by 900 cm$^{-1}$ exhibit in-plane hyperbolicity. The energy density within the IFC cone at 900 cm$^{-1}$ will be larger than that at 994 cm$^{-1}$. This can lead to stronger polariton reflection and transmission at the disk edge, which will result in longer decay length of the polariton field outside the disk. Second, the diameter of the disk excited at 900 cm$^{-1}$ (1.5 μm) is smaller than that excited by 994 cm$^{-1}$ (2.0 μm). A larger disk diameter will lead to a longer polariton propagation length and therefore a stronger damping. This will also make the polariton field decay fast outside the disk at 994 cm$^{-1}$.

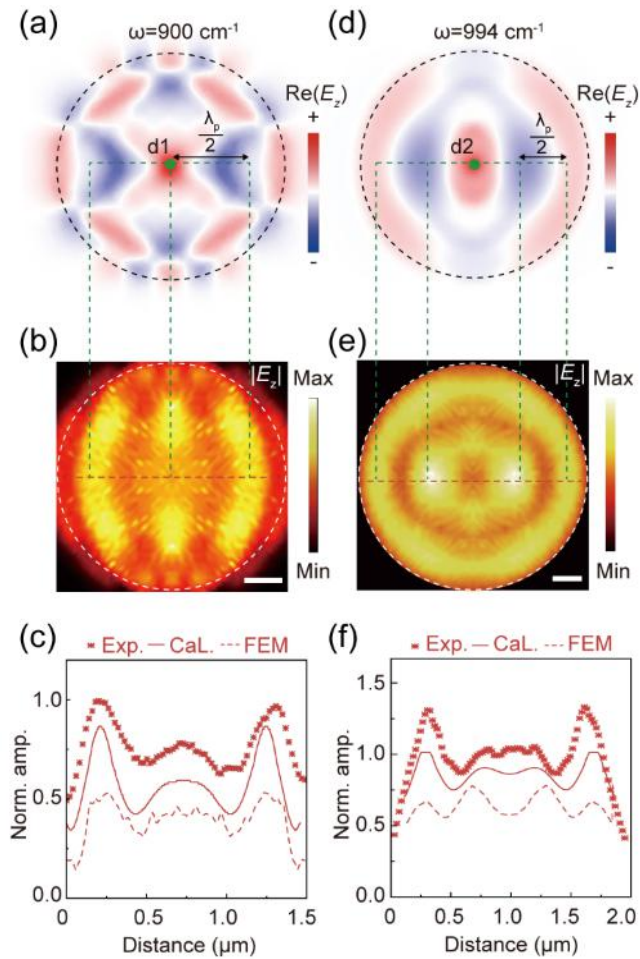

**Figure 3.** FEM simulations of HPhP interferences in a α-MoO$_3$ micro-disk. a, d) Simulated spatial near-field distributions of Re($E_z$). The z-polarized electric dipoles are placed at positions d1 and d2. b, e) Simulated near-field images ($|E_z|$). Scale bars: 250 nm. c, f) Experimental (symbols), simulated (dashed lines), and model calculated (solid lines) amplitude profiles extracted along the dashed lines marked in (b) and (e). The amplitudes are normalized by those from the substrate next to the disk. The excitation frequencies are 900 cm$^{-1}$ (a–c) and 994 cm$^{-1}$ (d–f).

Interference is governed by the phase factor $e^{iq\Delta}$, where $\Delta$ denotes the polariton propagation lengths that can be controlled by tuning the micro-disk size. We then perform nano-imaging on a set of micro-disks with varied radii ($L$ = 0.75, 1.0, and 1.26 μm) and fixed thicknesses ($t$ = 170 nm). Near-field images are recorded at two representative frequencies at 900 cm$^{-1}$ (negative-$\varepsilon_x$-Band) and 986 cm$^{-1}$ (elliptical band). Figure 4a−4c (Figure 4e−4g) depict that all of the micro-disks, irrespective of their sizes, exhibit highly anisotropic fringe shapes. The number of bright fringes particularly decreases as the disk size is reduced. These results can be validated by the corresponding cavity model calculations (Figure S9). To quantify the dependence of field localizations on the disk radius, we examine the near-field amplitudes at three different positions inside each disk as typical examples (indicated by letters "*a*," "*b*," and "*c*" in Figure 4a and 4e). For the two excitation frequencies, the experimental and calculated near-field amplitudes at these three exemplary positions all exhibit oscillations against $L$ (Figure 4d and 4h). In addition, a bright fringe periodically appears at the center of the disk with a decreasing $L$, providing clear evidence of the phase factor modulations. Another interesting observation is that the oscillation periods at *b* and *c* are different and can be seen more clearly from the model calculation results (solid lines in Figure 4d and 4h). For $\omega$ = 900 cm$^{-1}$, the period at *b* (452 nm) is smaller than that at *c* (458 nm), while for $\omega$ = 986 cm$^{-1}$, an opposite result is obtained (*i.e.*, the period at *b* (283 nm) is a bit larger than that at *c* (280 nm)). These behaviors can be understood by considering the phase factor $e^{iq\Delta}$ and the highly anisotropic in-plane polariton wavevectors in these two bands. In negative-$\varepsilon_x$-Band, the polariton interference at positions *b* and *c* are contributed by the polaritons with a large $q$ near the IFC asymptotes and a relatively small $q$ near the $q_x$-axis (Figure 2a). Accordingly, a smaller oscillation period against the disk size is observed at position *b*. In contrast, in elliptical band, the interference at position *b* is caused by the superposition of the polariton waves with smaller wavevectors (Figure 2b) giving rise to a larger oscillation period. These results clearly indicate that the micro-disk radius is an important parameter for configuring the spatial localization field distributions.

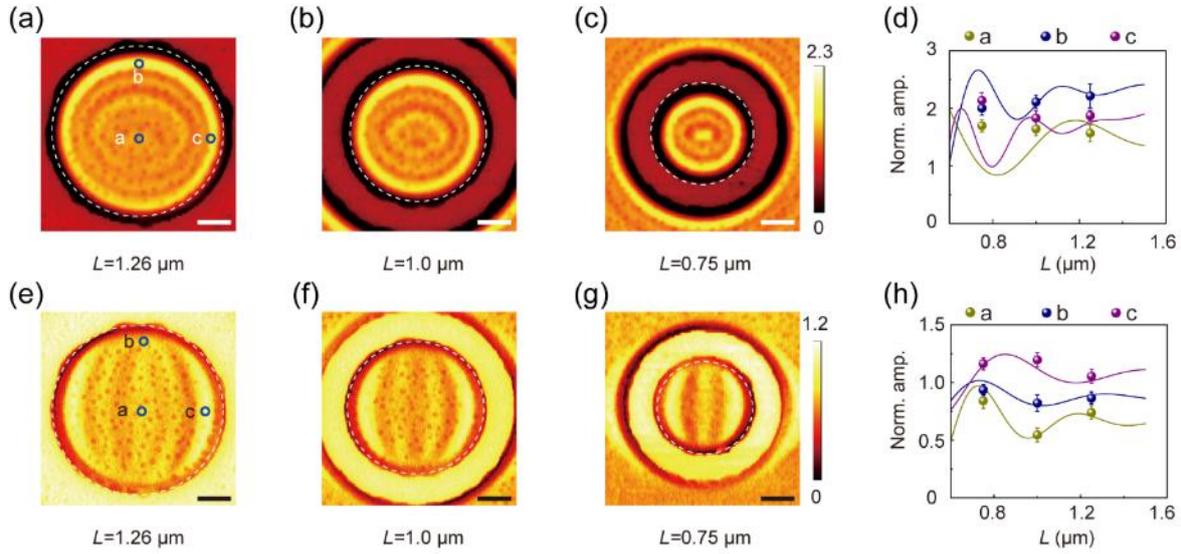

**Figure 4.** HPhP interferences in α-MoO$_3$ micro-disks of varied sizes. a–c and e–g) Experimental near-field amplitude images of micro-disks with different radii *L*. The excitation frequencies are 986 cm$^{-1}$ (a–c) and 900 cm$^{-1}$ (e–g). The white dashed circles indicate the disk boundaries. d, h) Normalized near-field amplitudes at positions *a* (green spheres), *b* (blue spheres), and *c* (purple spheres) marked in (a) and (e). Normalizations are performed by dividing the amplitudes in the micro-disk by that of the substrate. The green, blue, and purple solid lines are the corresponding calculations by the cavity model. The error bars in (d) and (h) are obtained based on the measurements on five points closest to *a*, *b*, and *c*. Scale bars: 500 nm.

As discussed before, the HPhP interferences are dominated by the boundaries perpendicular to the wavevectors of the polariton waves. Consequently, the interference fields can be tailored by modifying the shapes (*i.e.* boundary numbers and types) of the α-MoO$_3$ microstructures. Figure 5 illustrates the spatial near-field distributions obtained by the model calculations and the experimental measurements in three typical microstructures (Figure S7c, S7e, and S7f). For ease of discussion, the left most boundary of each microstructure is kept parallel to the *y*-axis ([001] crystalline direction). In contrast to the highly symmetric standing wave patterns formed by isotropic polariton superpositions,[33, 39] the in-plane anisotropic HPhPs will lead to abnormal asymmetric interference patterns in each microstructure.

Specifically, for excitations in elliptical band (986 cm$^{-1}$), where the HPhPs can spread along all directions with orientation-dependent wavevectors, all boundaries can act as reflectors and generate polariton interferences that are strongly dependent on the microstructure shapes. For square (Figure 5a and 5g), regular pentagon (Figure 5b and 5h), and regular hexagon (Figure 5c and 5i), the bright spots formed by constructive interferences generally exhibit spatial distributions distorted from the geometrical symmetries of the microstructures. This can be more directly observed on the near-field profiles crossing the microstructure centers along the *x*- and *y*-axes (solid and dashed lines, respectively, in Figure 5m−5o). In the square and regular hexagonal, the bright spots particularly follow elongated distributions along the *x*-axis ([100] crystalline direction). Thus, the fringe spacings along the *x*-axis are smaller than those along the *y*-axis (solid and dashed lines, respectively, in Figure 5m, o) due to the larger wavevector of the HPhPs parallel to the *x*-axis (Figure 2b).

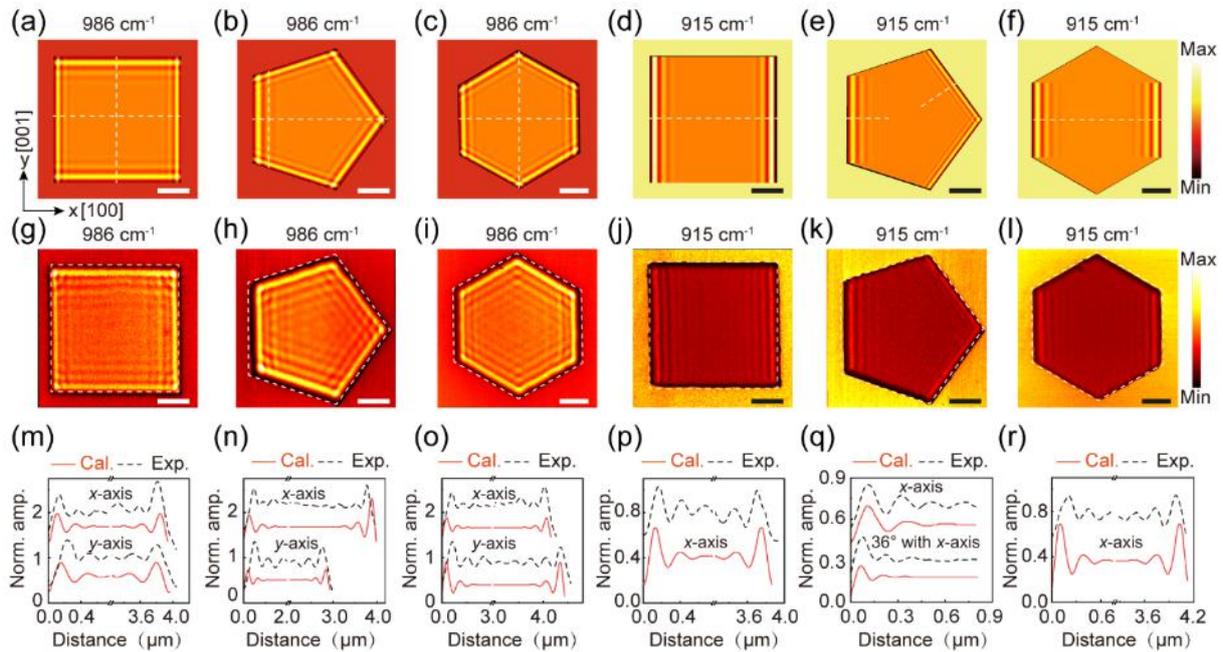

**Figure 5.** HPhP interferences in α-MoO$_3$ microstructures of varied shapes. a−c) Calculated interference patterns of square (a), regular pentagon (b), and regular hexagon (c) microstructures. The excitation frequencies are 986 cm$^{-1}$. d−f) Corresponding calculated interference patterns using excitations at 915 cm$^{-1}$. g−l) Experimental near-field amplitude images of the microstructures corresponding to those in (a−f). Dashed white lines mark the boundaries. m−r) Normalized field profiles along the dashed white lines indicated in (a−f). The profiles for (e) and (k) are drawn along *x*-axis and direction of 36° with respect to *x*-axis.

The normalizations are done by dividing the calculated or experimental field amplitudes of the microstructures by that of the substrate. Scale bars: 1.0 μm.

The interference patterns become even more asymmetrical for the excitation frequency in negative-$\varepsilon_x$-Band (915 cm$^{-1}$). Both the square and regular hexagon microstructures only exhibit fringes parallel to the *y*-axis, with the two strongest located next to the edges (Figure 5d, 5j, 5f, and 5l). This result can be understood by considering that the IFC at 915 cm$^{-1}$ is a hyperbola opening toward the *x*-axis. The opening angle between the two asymptotes is $\varphi = 108°$. Therefore, for polaritons with wavevectors confined by the two asymptotes, only the two boundaries parallel to the *y*-axis can reflect the polariton waves, leading to the observed fringes along the *y*-axis. The polariton wavelengths extracted from the near-field profiles of these two microstructures are 475 nm and 480 nm (Figure 5p and 5r), respectively. These values match well with $\lambda_p = 2\pi/q_{[100]} = 480$ nm, where $q_{[100]}$ is the wavevector determined from the IFC for the HPhPs propagating along the *x*-axis. However, for a regular pentagon microstructure, the angle between the two edges intersecting with the *x*-axis is 108°. Two polariton waves can always be back-reflected by these two edges because $\pi - 108° = 72° < \varphi$. As a result, additional fringes parallel to the two edges can be observed (Figure 5e and 5k). Moreover, due to the forbidden of polariton propagation, no fringes emanating from the rotated edges can be observed in all of the three microstructures (Figure 5d−5f and 5j−5l). Most interestingly, according to the IFC, the *q* associated with the polariton waves propagating toward the two oblique edges is much larger than $q_{[100]}$ (Figure 2a), thereby rendering much shorter polariton wavelengths ($\lambda_p = 190$ nm) (Figure 5q) and indicating much stronger electric field confinements near the oblique edges. Another feature should be noted is that slightly deformed fringes can be observed experimentally at corners of square microstructure, while they are absent in the calculated image (Figure 5d and 5j). These deformed fringes are due to interference involve high-wavevector polaritons, which are similar to those observed in the circular micro-disk (Figure 2c−2j). As mentioned before, in our calculations only the boundary perpendicular to the polariton wavevector is considered. Therefore, in the square microstructure, the contributions from polaritons with high wavevectors are neglected, giving rise to discrepancy of interference fringes at the corners between calculation and experimental results.

The results above unambiguously demonstrate that the electromagnetic field localizations can be configured by controlling the shape and the size of an α-MoO$_3$ microstructure, and the excitation frequency. We further demonstrate the configurability by fabricating and imaging a set of α-MoO$_3$ micro-wedges with fixed vertex angles at 30° and varied skew angles ($β$) from 0 to 90°, which is defined by the angle between the wedge bisector and the *y*-axis (upper panel, Figure 6a). The near-field distributions were first examined at an excitation frequency of 990 cm$^{−1}$, where the IFC is an ellipse (Figure 2b). Two polariton waves that can be back-reflected by the two edges always exist, leading to the interference fringes parallel to the wedge boundaries (Figure 6a). In particular, the orientations of the two edges relative to the major axes of the IFC will be adjusted by rotating the wedge around the *z*-axis (*i.e.*, changing the $β$). Therefore, the two reflected polariton waves will exhibit different wavevectors, enabling modifications of their interference fields (see note S4 and Figure S10). For ease of discussion, the edge above (below) the *x*-axis is labeled as Edge I (Edge II) (as indicated in Figure 6a). For $β$ = 0 and 90°, the wedges exhibit lateral symmetries relative to the *y*- and *x*-axes, respectively. The polariton waves perpendicularly reflected by the two edges exhibit the same *q* (Figure S10), making the two sets of fringes close to the edges equivalent (Figure S11). The *q* of the polariton wave reflected by Edge I will first be reduced and then increased as $β$ is steadily decreased from 90°. This is opposite to that of the polariton wave reflected by Edge II (Figure S11). Consequently, the corresponding two sets of fringes will differ from each other in terms of their spacings and linewidths (Figure S11). Specifically, $λ_p$ associated with Edge I (Edge II) will first increase (decrease) and then decrease (increase), as confirmed by the experimental measurements (Figure 6b). The model calculation results agree very well with the experimental measurements.

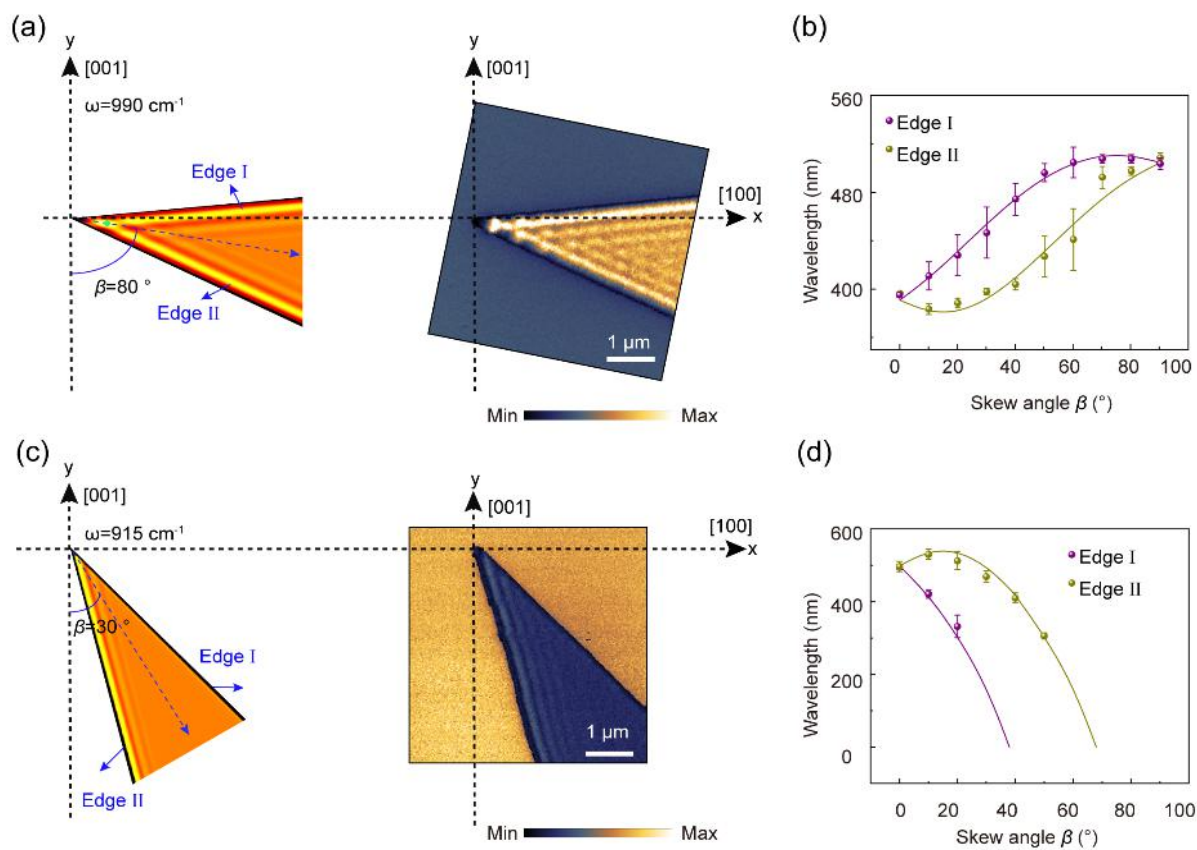

**Figure 6.** HPhP interference effects in α-MoO$_3$ micro-wedges of varied skew angles and fixed vertex angles. a, c) Calculated and experimental HPhP interference patterns in a representative micro-wedge. Left panels: schemes of typical micro-wedges. The skew angle is defined by the angle between the bisector of the wedge and the *y*-axis. The two edges are labeled as Edge I and Edge II, respectively. Right panels: typical near-field image of the micro-edge. The excitation frequency is 990 cm$^{-1}$ (elliptical band). b, d) Dependence of the HPhP wavelength on the skew angle of the wedge. The wavelengths are extracted from the fringes parallel to Edge I and Edge II. The solid lines and symbols in (b) and (d) denote the calculation and experimental results, respectively. Error bars in (b) and (d) are based on analyses of five typical amplitude line profiles perpendicular to Edge I and Edge II, respectively, in each experimental near-field image. The micro-wedge thickness is 170 nm. The excitation frequencies are 990 cm$^{-1}$ (a and b) and 915 cm$^{-1}$ (c and d). Scale bars: 1.0 μm.

The interference patterns are more interesting for an excitation frequency of 915 cm$^{-1}$ in the negative-$\varepsilon_x$-Band. For $\beta$ = 90°, because the angle of the two wedge edge normal directions is 150°, polariton waves cannot propagate perpendicularly towards Edge I and II (Figure S12a). No interference fringes can be observed parallel to the two edges (Figure S13a). As $\beta$ is progressively decreased, the two normal directions will successively sweep the two asymptotes of the IFC (Figure S12a and S12b), enabling polariton reflection and interference at the edges. Specifically, when the $β$ is reduced by more than 21° ($β < 69°$), the normal direction of Edge II will intersect with the IFC, while the normal direction of Edge I is still outside the IFC. As a result, polariton fringes parallel to Edge II appear, while those parallel to Edge I are still absent (Figure S13a). When $β$ is further reduced, polariton wave vector $q_{II}$ will be reduced and then increased again as $β$ is smaller than 15° (Figure S12b and S12c). Accordingly, polariton wavelength $λ_p$ associated with Edge II will first increase and then decrease against $β$ (Figure 6d and Figure S13a). On the other hand, the normal direction of Edge I can intersect with the IFC only when $β$ is smaller than 39°. Therefore, the fringes parallel to Edge I will not appear unless $β < 39°$ (Figure 6c and Figure S12c). Afterwards, the wave vector $q_I$ is reduced against $β$, giving rise to monotonically increased $λ_p$ at Edge I (Figure 6d). These experimental findings are well corroborated with the model calculations (Figure 6d and Figure S13b).

## 3. Conclusion

The patterning of photonic materials into micro- and nanostructures has been well established for controlling electromagnetic waves at the nanoscale. Our study demonstrates that by combining in-plane anisotropic polaritons with high spatial confinement, that polariton interference effects inside an α-MoO$_3$ microstructures can be controlled by tuning the excitation frequency (IFC topology), shape (number and type of the boundary), and size (polariton propagation length) of the microstructure. Thus, the configurability of the nanoscale electromagnetic fields can be extended in terms of anisotropic spatial distributions and orientation dependence, which, otherwise, usually requires the design and patterning of complex architectures in isotropic counterparts. It should be noted that very recently two in-plane THz polaritonic bands have been discovered in α-MoO$_3$.[16] The results obtained in the current study can therefore also be applied to the THz spectral region, which can help design of nanophotonic devices for THz applications. In principle, this rationale can also be generalized to other types of vdW crystals and artificial metasurfaces with in-plane optical responses, which is expected to open up a new paradigm for confining and manipulating light flow in planar photonics. From a fundamental point of view, the α-$MoO_3$ microstructures sustaining highly anisotropic localized electromagnetic fields can provide a testing platform for studying light−matter interactions in complex spatially confined electromagnetic environments.

During submission of our manuscript, we became aware of a very recent publication reporting similar studies.[40] Our work was conducted independently. In addition, in our present study, we further demonstrated configuring the electromagnetic field localizations in α-$MoO_3$ microstructures of different sizes and shapes.

## 4. Experimental Section

*Fabrication of α-$MoO_3$ Microstructures*: α-$MoO_3$ single crystal slabs were prepared using thermal physical vapor deposition method.[41] The crystals were directly synthesized on silicon substrate grown with 300-nm thick $SiO_2$ layer. The various α-$MoO_3$ microstructures were fabricated using the focused ion beam (FIB) technique. Specifically, $Ga^+$ ions were used as ion sources in our FIB etching system (AURIGA, Zeiss). The acceleration voltage and current of $Ga^{3+}$ beam were respectively set as 30 kV and 10 pA, with a $Ga^{3+}$ beam dose of 1 nC/$\mu m^2$. The dwell time was 0.5 μs. After the microstructures were constructed, they were annealed at 300 °C for 2 h in $O_2$ ambient condition to eliminate the $Ga^{3+}$ embedded in the α-$MoO_3$.

*Numerical Simulations*: Numerical simulations were performed using the finite element method (FEM, Comsol). To generate the near-field spatial distributions above the various 2D slabs and microstructures, HPhPs were launched using a *z*-polarized electric dipole source.

Specifically, in each simulation, the dipole was located 200-nm above the sample surface. The near-field distributions, Re($E_z$), were obtained on the plane 20-nm above the sample surface. The dipole was scanned across the sample at a step of 5 nm. The thicknesses of the slabs and α-MoO$_3$ microstructures were set as 170 nm. Permittivities of the samples were modeled by fitting their respective experimental data using Lorentzian dielectric models.[3, 15]

*Nanoimaging of Various α-MoO$_3$ Microstructures*: Real-space nanoimaging was performed using an s-SNOM (NeaSNOM, Neaspec GmbH), which was built based on an atomic force microscope (AFM). In a specific measurement, the metal-coated tip (Arrow-IrPt, NanoWorld) was illuminated by a mid-infrared laser source with tunable frequencies from 900 to 1240 cm$^{-1}$ (quantum cascade laser, Daylight Solutions). The tip was vibrated at a frequency of ~270 kHz. The back-scattered light from the tip was directed to an MCT detector (HgCdTe, Kolmar Technologies). The near-field signal was then extracted using a pseudo-heterodyne interferometric method, and the detected signal was demodulated at a third harmonic of the tip vibration frequency.

*Statistical Analysis*: The error bars in Figure 2l and Figure 2n are obtained based on the two symmetrical points along the *y*-axis on the outmost fringes in Figure 2h and Figure 2j. To obtain the error bars in Figure 4d and Figure 4h, five points closest to positions *a*, *b*, and *c* in Figure 4a and Figure 4e are analyzed. Error bars in Figure 6b and 6d are based on analyses of five typical amplitude line profiles perpendicular to Edge I and Edge II, respectively, in each experimental near-field image. All of the data were represented in form of average ± standard deviations.

**Supporting Information**
Supporting Information is available from the Wiley Online Library or from the author.

**Acknowledgements**
W.H., F.S., and Z. Z. contributed equally to this work. We acknowledge support from the National Key Basic Research Program of China (grant no. 2019YFA0210203), the National

Natural Science Foundation of China (grant nos. 91963205 and 11904420), Guangdong Basic and Applied Basic Research Foundation (grant nos. 2019A1515011355 and 2020A1515011329). H.C. acknowledges the support from Changjiang Young Scholar Program. Z.Z. acknowledges the project funded by China Postdoctoral Science Foundation (grant no. 2019M663199). J.D.C. acknowledges support from the National Science Foundation, Division of Materials Research (grant no. 1904793), while support for T.G.F. was provided by Vanderbilt University through the startup package of J.D.C..

**Conflict of Interest**
The authors declare no conflict of interest.

Received: ((will be filled in by the editorial staff))
Revised: ((will be filled in by the editorial staff))
Published online: ((will be filled in by the editorial staff))

References

[1] R. Hillenbrand, T. Taubner, F. Keilmann, *Nature* **2002**, *418*, 159−162.

[2] T. Dekorsy, V. A. Yakovlev, W. Seidel, M. Helm, F. Keilmann, *Phys. Rev. Lett.* **2003**, *90*, 055508.

[3] J. D. Caldwell, A. V. Kretinin, Y. G. Chen, V. Giannini, M. M. Fogler, Y. Francescato, C. T. Ellis, J. G. Tischler, C. R. Woods, A. J. Giles, M. Hong, K. Watanabe, T. Taniguchi, S. A. Maier, K. S. Novoselov, *Nat. Commun.* **2014**, *5*, 5221.

[4] S. Dai, Z. Fei, Q. Ma, A. S. Rodin, M. Wagner, A. S. Mcleod, M. K. Liu, W. Gannett, W. Regan, K. Watanabe, T. Taniguchi, M. Thiemens, G. Dominguez, A. H. C. Neto, A. Zettl, F. Keilmann, P. Jarillo-Herrero, M. M. Fogler, D. N. Basov, *Science* **2014**, *343*, 1125–1129.

[5] J. D. Caldwell, L. Lindsay, V. Giannini, I. Vurgaftman, T. L. Reinecke, S. A. Maier, O. J. Glembocki, *Nanophotonics* **2015**, *4*, 44–68.

[6] T. Low, A. Chaves, J. D. Caldwell, A. Kumar, N. X. Fang, P. Avouris, T. F. Heinz, F. Guinea, L. Martin-Moreno, F. Koppens, *Nat. Mater.* **2017**, *16*, 182–194.

[7] D. N. Basov, M. M. Fogler, F. J. García de Abajo, *Science* **2016**, *354*, aag1992.

[8] W. Ma, P. Alonso-González, S. Li, A. Y. Nikitin, J. Yuan, J. Martín-Sánchez, J. Taboada-Gutiérrez, I. Amenabar, P. Li, S. Vélez, C. Tollan, Z. Dai, Y. Zhang, S. Sriram, K. Kalantar-Zadeh, S. -T. Lee, R. Hillenbrand, Q. Bao, *Nature* **2018**, *562*, 557–562.

[9] Z. Zheng, N. Xu, S. L. Oscurato, M. Tamagnone, F. Sun, Y. Jiang, Y. Ke, J. Chen, W. Huang, W. L. William, A. Ambrosio, S. Deng, H. Chen, *Sci. Adv.* **2019**, *5*, eaav8690.

[10] S. Dai, Q. Ma, T. Andersen, A. S. Mcleod, Z. Fei, M. K. Liu, M. Wagner, K. Watanabe, T. Taniguchi, M. Thiemens, F. Keilmann, P. Jarillo-Herrero, M. M. Fogler, D. N. Basov, *Nat. Commun.* **2015**, *6*, 6963.

[11] P. Li, M. Lewin, A. V. Kretinin, J. D. Caldwell, K. S. Novoselov, T. Taniguchi, K. Watanabe, F. Gaussmann, T. Taubner, *Nat. Commun.* **2015**, *6*, 7507.

[12] M. Autore, P. Li, I. Dolado, F. J. Alfaro-Mozaz, R. Esteban, A. Atxabal, F. Casanova, L. E. Hueso, P. Alonso-González, J. Aizpurua, A. Y. Nikitin, S. Vélez, R. Hillenbrand, *Light Sci. Appl.* **2018**, *7*, 17172.

[13] S. -A. Biehs, M. Tschikin, P. Ben-Abdallah, *Phys. Rev. Lett.* **2012**, *109*, 104301.

[14] D. G. Baranov, Y. Xiao, I. A. Nechepurenko, A. Krasnok, A. Alù, M. A. Kats, *Nat. Mater.* **2019**, *18*, 920–930.

[15] Z. Zheng, J. Chen, Y. Wang, X. Wang, X. Chen, P. Liu, J. Xu, W. Xie, H. Chen, S. Deng, N. Xu, *Adv. Mater.* **2018**, *30*, 1705318.

[16] T. V. A. G. D. Oliveira, T. Nörenberg, G. Álvarez-Pérez, L. Wehmeier, J. Taboada-Gutiérrez, M. Obst, F. Hempel, E. J. H. Lee, J. M. Klopf, I. Errea, A. Y. Nikitin, S. C. Kehr, P. Alonso-González, L. M. Eng, *Adv. Mater.* **2021**, *33*, 2005777.

[17] J. Taboada-Gutiérrez, G. Álvarez-Pérez, J. Duan, W. Ma, K. Crowley, I. Prieto, A. Bylinkin, M. Autore, H. Volkova, K. Kimura, T. Kimura, M. -H. Berger, S. Li, Q. Bao, X. P. A. Gao, I. Errea, A. Y. Nikitin, R. Hillenbrand, J. Martín-Sánchez, P. Alonso-González, *Nat. Mater.* **2020**, *19*, 964–968.

[18]  G. Hu, Q. Ou, G. Si, Y. Wu, J. Wu, Z. Dai, A. Krasnok, Y. Mazor, Q. Zhang, Q. Bao, C. -W. Qiu, A. Alù, *Nature* **2020**, *582*, 209–213.

[19]  M. Chen, X. Lin, T. H. Dinh, Z. Zheng, J. Shen, Q. Ma, H. Chen, P. Jarrillo-Herrero, S. Dai, *Nat. Mater.* **2020**, *19*, 1307–1311.

[20]  Z. Zheng, F. Sun, W. Huang, J. Jiang, R. Zhan, Y. Ke, H. Chen, S. Deng, *Nano Lett.* **2020**, *20*, 5301–5308.

[21]  J. Duan, N. Capote-Robayna, J. Taboada-Gutiérrez, G. Álvarez-Pérez, I. Prieto, J. Martín-Sánchez, A. Y. Nikitin. P. Alonso-González, *Nano Lett.* **2020**, *20*, 5323–5329.

[22]  S. Dai, Q. Ma, M. K. Liu, T. Andersen, Z. Fei, M. D. Goldflam, M. Wagner, K. Watanabe, T. Taniguchi, M. Thiemens, F. Keilmann, G. C. A. M. Janssen, S. -E. Zhu, P. Jarillo-Herrero, M. M. Fogler, D. N. Basov, *Nat. Nanotechnol.* **2015**, *10*, 682–686.

[23]  J. Duan, R. Chen, R, J. Li, K. Jin, Z. Sun, J. Chen, *Adv. Mater.* **2017**, *29*, 1702494.

[24]  T. G. Folland, A. Fali, S. T. White, J. R. Matson, S. Liu, N. A. Aghamiri, J. H. Edgar, R. F. Haglund Jr, Y. Abate, J. D. Caldwell, *Nat. Commun.* **2018**, *9*, 4371.

[25]  A. M. Dubrovkin, B. Qiang, H. N. S. Krishnamoorthy, N. I. Zheludev, Q. Wang, *Nat. Commun.* **2018**, *9*, 1762.

[26]  A. D. Dunkelberger, C. T. Ellis, D. C. Ratchford, A. J. Giles, M. Kim, C. S. Kim, B. T. Spann, I. Vurgaftman, J. G. Tischler, J. P. Long, O. J. Glembocki, J. C. Owrutsky, J. D. Caldwell, *Nat. Photonics* **2018**, *12*, 50–56.

[27]  H. Karakachian, M. Kazan, *J. Appl. Phys.* **2017**, *121*, 093103.

[28]  A. J. Giles, S. Dai, S, I. Vurgaftman, T. Hoffman, S. Liu, L. Lindsay, C. T. Ellis, N. Assefa, I. Chatzakis, T. L. Reinecke, J. G. Tischler, M. M. Fogler, J. H. Edgar, D. N. Basov, J. D. Caldwell, *Nat. Mater.* **2018**, *17*, 134–139.

[29]  Y. Wu, Q. Ou, Y. Yin, Y. Li, W. Ma, W. Yu, G. Liu, X. Cui, X. Bao, J. Duan, G. Álvarez-Pérez, Z. Dai, B. Shabbir, N. Medhekar, X. Li, C. -M. Li, P. Alonso-González, Q. Bao, *Nat. Commun.* **2020**, *11*, 2646.

[30] S. Dai, M. Tymchenko, Y. Yang, Q. Ma, M. Pita-Vidal, K. Watanabe, T. Taniguchi, P. Jarillo-Herrero, M. M. Fogler, A. Alù, D. N. Basov, *Adv. Mater.* **2018**, *30*, 1706358.

[31] B. Liao, X. Guo, D. Hu, F. Zhai, H. Hu, K. Chen, C. Luo, M. Liu, X. Xiao, Q. Dai, *Adv. Funct. Mater.* **2019**, *29*, 1904662.

[32] K. Chaudhary, M. Tamagnone, M. Rezaee, D. K. Bediako, A. Ambrosio, P. Kim, F. Capasso, *Sci. Adv.* **2019**, *5*, eaau7171.

[33] Z. Zheng, J. Li, T. Ma, H. Fang, W. Ren, J. Chen, J. She, Y. Zhang, F. Liu, H. Chen, S. Deng, N. Xu, *Light Sci. Appl.* **2017**, *6*, e17057.

[34] N. Ocelic, R. Hillenbrand, *Nat. Mater.* **2004**, *3*, 606–609.

[35] A. M. Dubrovkin, B. Qiang, T. Salim, D. Nam, N. I. Zheludev, Q. Wang, *Nat. Commun.* **2020**, *11*, 1863.

[36] F. Sun, W. Huang, Z. Zheng, N. Xu, Y. Ke, R. Zhan, H. Chen, S. Deng, *Nanoscale*, DOI:10.1039/D0NR07372E, **2021**.

[37] G. Álvarez-Pérez, K. V. Voronin, V. S. Volkov, P. Alonso-González, A. Y. Nikitin, *Phys. Rev. B* **2019**, *100*, 235408.

[38] J. A. Gerber, S. Berweger, B. T. O'Callahan, M. B. Raschke, *Phys. Rev. Lett.* **2014**, *113*, 055502.

[39] A. Y. Nikitin, P. Alonso-González, S. Vélez, S. Mastel, A. Centeno, A. Pesquera, A. Zurutuza, F. Casanova, L. E. Hueso, F. H. L. Koppens, R. Hillenbrand, *Nat. Photonics* **2016**, *10*, 239–243.

[40] Z. G. Dai, G. W. Hu, G. Y. Si, Q. D. Ou, Q. Zhang, S. Balendhran, F. Rahman, B. Y. Zhang, J. Z. Ou, G. G. Li, A. Alù, C.-W. Qiu, Q. L. Bao, *Nat. Commun.* **2020**, *11*, 6086.

[41] Y. Wang, X. Du, J. Wang, M. Su, X. Wan, H. Meng, W. Xie, J. Xu, P. Liu, *ACS Appl. Mater. Interfaces* **2017**, *9*, 5543–5549.

*Wuchao Huang, Fengsheng Sun, Zebo Zheng, Thomas G. Folland, Xuexian Chen, Huizhen Liao, Ningsheng Xu, Joshua D. Caldwell, Huanjun Chen\* & Shaozhi Deng\**

**Title** *Van der Waals phonon polariton microstructures for configurable infrared electromagnetic field localizations*

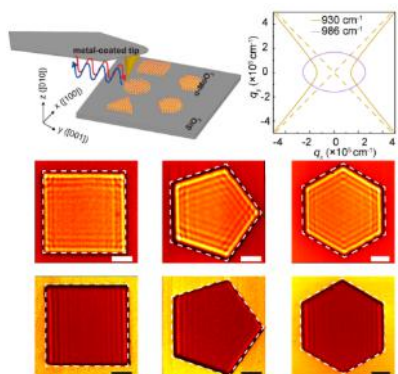

By taking advantage of the in-plane hyperbolic phonon polariton and spatial confinement effects, we demonstrate the nanoscale tailoring of mid-infrared electromagnetic field localizations in van der Waals α-$MoO_3$ microstructures of different shapes and sizes. We further show that orientation of the in-plane isofrequency curve relative to the microstructure edge is a critical parameter governing the localized electromagnetic field distributions.